# PID2018 Benchmark Challenge:
# Multi-Objective Stochastic Optimization Algorithm

**Abdullah Ates***, **Jie Yuan****, **Sina Dehghan*****, **Yang Zhao******,
**Celaleddin Yeroglu***, **YangQuan Chen*****

*Inonu University, Computer Engineering Department, 44280, Malatya, TURKEY
( e-mail: abdullah.ates@inonu.edu.tr, c.yeroglu@inonu.edu.tr ).
** School of Automation, Southeast University, Nanjing, 210096, CHINA
(e-mail: 230149417@seu.edu.cn )
*** UC Merced, Mechanical Engineering Departments, MESA Lab, Merced, CA,95301, USA,
(e-mail: ychen53@ucmerced.edu , sdehghan@ucmerced.edu,)
**** School of Control Science and Engineering at Shandong University, Jinan, 250061 CHINA
(e-mail: zdh1136@gmail.com )

**Abstract:** This paper presents a multi-objective stochastic optimization method for tuning of the controller parameters of Refrigeration Systems based on Vapour Compression. Stochastic Multi Parameter Divergence Optimization (SMDO) algorithm is modified for minimization of the Multi Objective function for optimization process. System control performance is improved by tuning of the PI controller parameters according to discrete time model of the refrigeration system with multi objective function by adding conditional integral structure that is preferred to reduce the steady state error of the system. Simulations are compared with existing results via many graphical and numerical solutions.

1. INTRODUCTION

Many different model structures were used for refrigeration systems. Vapour-compression refrigeration systems are widely used for domestic, commercial and industrial refrigeration because of its big advantage of saving energy (Bejarano et al. (2018)).

Thus, control of the refrigeration system is a very challenging area for the researchers. In the literature, many studies were discussed the issue due to potential of future usage. Decentralized systems PID control (Bejarano et al. (2018); Marcinichen et al. (2008); Salazar and Méndez (2014)), decoupling multivariable control (Shen et al. (2010)), LQG control (Schurt et al. (2009-2010), model predictive control (MPC) (Razi et al. (2006)), and robust H∞ control (Bejarano et al., 2015)) are some of the prominent studies in this area.

PID and PI controllers, known as the most effective control structures for many decades, give also satisfactory result for the Vapour-compression refrigeration system. Furthermore, usage of the optimized controller parameters can further improve the control performance.

Generally, optimization can be classified in two groups that are based on analytical methods and numerical methods. Analytical methods have high computational complexity and hard to implement on complicated system like the model of vapour-compression refrigeration systems. But numerical methods particularly stochastic optimization methods have many advantage for decrease of the mathematical complexity. There are many studies can be found about optimization of the controller parameters with stochastic and heuristic methods such as SMDO method (Alagoz et al. (2013); Yeroglu and Ates (2014)), Tabu search based optimization algorithm (Ates and Yeroglu (2016)), Fruit Fly Optimization algorithm (Sheng and Yan, (2013), Cuckoo search algorithm (Zamani et al., (2017)), adaptive particle swarm optimization algorithm (Liu, (2016)).

Usage of the multi objective function in the optimization process is another advantage of the stochastic optimization methods in control applications. Especially, controller tuning problem can give satisfactory results with multi objectives for fast settling time, low overshoot, less steady state error, etc. Furthermore, multi objective functions can be based on the error functions that are MSE, ITAE, ISE and IAE.

In the literature many studies deal with multi objective optimization. For instance, FOPID controller design with Big Bang Big Crunch optimization algorithm with multi objective function is proposed in (Ates et al. (2017)). Tuning of PI and PID controller parameters by using overshoot, steady state error, rise time and settling time is presented in (Zeng et al. (2015)). FOPID controller parameters are optimized with particle swarm optimization algorithm according to multi objective function that are settling time, overshoot, integral of the square input, steady state error, rising time, IAE, gain margin and phase margin (Zamani et al. 2009)). Adaptive grid particle swarm optimization is used to optimize the PID gains tuning problem of the hydraulic turbine regulating system with multi objective function (Chen et al., (2015)). A genetic algorithm proposes for tuning of the PID controller according to multi objective function that is combined with rise time, settling time, percentage over-shoot and integral of error squared (Neath et al. (2014)). Optimization of the PID controller based on multi objective optimization and genetic algorithms is proposed in (Herreros et al. (2002)). Two PID controllers' parameters are optimized with Non-dominated sorting genetic Algorithm approach according to multi objective function for two degree of freedom robot manipulator in (Ayala and Coelho, (2012)). A novel super-twisting PID sliding mode controller is proposed by using a multi-objective optimization bat algorithm for the control of gyroscope in (Jayabarathi et al. (2018)).

In this paper, we proposed to extend SMDO algorithm for minimization of the multi objective function and presented Multi Objective SMDO (MO-SMDO) for tuning of PI controller parameters of Refrigeration Systems based on Vapour Compression. The proposed MO-SMDO can minimize two outputs of the corresponding system's ($T_{e,\text{sec},out}$ and $TSH$) error simultaneously.

Furthermore, conditional integration (Garcia and Castelo, 2002)) is used to reduce of the steady state error. Combination of the MO-SMDO and conditional integral structure can further improve the system control performance. These improvements are validated with some graphical and numerical results.

## 2. CONTROL STRUCTURE OF REFRIGERATOR SYSTEMS

Fig. 1 presents canonical one-compression-stage, one-load-demand refrigeration cycle. Fundamental components of the system that are expansion valve, compressor, evaporator and condenser are illustrated in the Fig. 1. Many information for working strategy of the refrigerator system and model identification of the corresponding systems can be found in (Bejarano et al. (2018)). In this study, only improving of the control structure is presented. For this reason, brief information is given only for the Refrigeration Systems based on Vapour Compression.

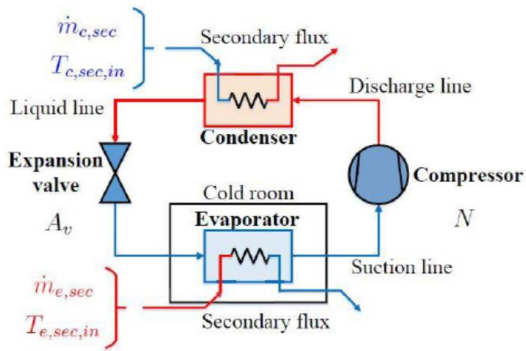

Fig.1. Schematic picture of one-compression-stage, one-load-demand vapour-compression refrigeration cycle (Bejarano et al. (2018)).

This system is controlled according to two inputs and two outputs. System inputs are evaporator secondary flux and $T_{e,\text{sec},out}$ and the degree of superheating $TSH$. The main control objective of the system is cooling power (Bejarano et al. (2018)).

Furthermore, system has two controllers that are discrete transfer function for $T_{e,\text{sec},out}$ in Fig 2 and discrete PI controller for $TSH$ in Fig. 3. Also Fig. 4 shows Matlab simulation model of the the Refrigeration Systems based on Vapour Compression.

In this structure optimization of the discrete PI controller parameters with stochastic optimization method according to multi objective function can improve the control performance.

Default system does not have conditional integral structure. Moreover, this system has some steady state error. It is not possible to eliminate steady state error with only tuning controller parameters. However, steady state error can be eliminated with usage of the conditional integral (CI) (Garcia and Castelo, 2002)).

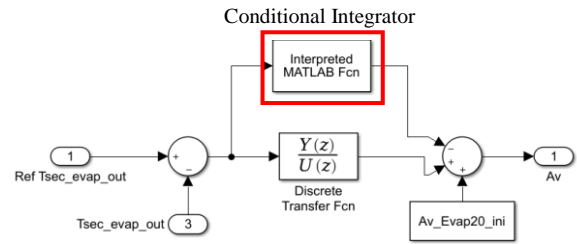

Fig.2. Control Structure of the $T_{e,\text{sec},out}$ (Bejarano et al. (2018)).

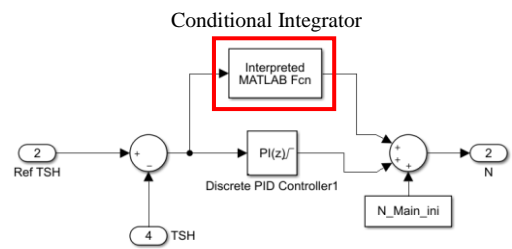

Fig.3. Control Structure of the $TSH$ (Bejarano et al. (2018)).

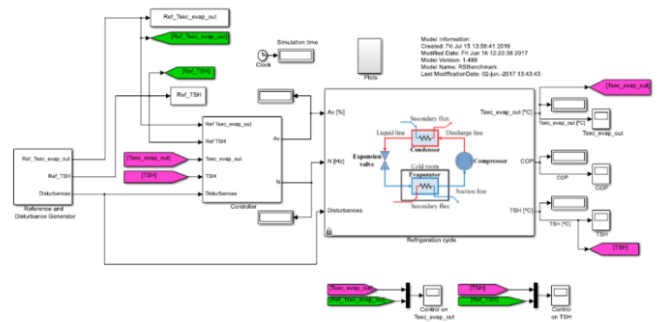

Fig.4. Matlab Simulation model of the Refrigeration Systems based on Vapour Compression.

## 3. MULTI OBJECTIVE STOCHASTIC MULTI PARAMETER DIVERGENCE OPTIMIZATION ALGORITHM

SMDO method is proposed and used for controller tuning problems in (Yeroglu and Ates (2014); Alagoz et al. (2013)). Especially SMDO is used for controller tuning problem without model identification. If the system mathematical model is not known exactly, SMDO method can optimize the controller or system parameter by minimizing the error function. Error function is an objective function for controller parameters optimization. Many studies can be found in the literature about optimization of the controller parameters with SMDO method (Alagoz et al., (2013); Yeroglu and Ates (2014); Ates et al. (2014); Ates et al. (2016); Ates et al. (2017)).

SMDO method search the parameter vector $P^n$ in the parameters work space randomly with a random movement in two direction that are forward direction, $\Delta P^n = \Psi_v^n \xi$ and backward direction, $\Delta P^n = -\Psi_v^n \xi$ respectively. For this way optimization algorithm can escape the global optimum. $\Psi_v^n$ is a parameter vector for corresponding parameters, $\xi$ is random weight function that can alter in between [0,1]. Sub-index $v$ denotes the component of $P$ vector. Parameters vectors for the tuning of the PI parameters is given as $P^n = [k_p \ k_i]$. During optimization, the cost function is computed for each test. Because SMDO method can search the parameters in the parameter search space according to cost function value (Alagoz et al., (2013); Yeroglu and Ates (2014); Ates et al. (2014); Ates et al. (2016); Ates et al. (2017)).

In this study PI controller parameters that are used in the control structure in Fig. 4 are optimized. This structure has two outputs that are $TSH$ and $T_{e,\text{sec},out}$. The structure and parameters should be enhanced for the two outputs simultaneously. Usage of one error function is not adequate for improving to both system output. For this reason, the cost function formulation is defined as a multi objective function by using $TSH$ and $T_{e,\text{sec},out}$ values. This multi objective formulation is given as follows:

$$E = \min(w_1 E_{TSH} + w_2 E_{e,\text{sec},out}) \quad (1)$$

where $E_{TSH}$ and $E_{T\text{sec}}$ error function values for $TSH$ and $T_{e,\text{sec},out}$ outputs respectively. These values are obtained by comparing between reference signals with generated output. $w_1$ and $w_2$ are random weighting functions that are changed between [0,1] for $E_{TSH}$ and $E_{T\text{sec}}$ respectively.

First, SMDO method starts searching with forward test according to following formulation

$$E(P^n + \Psi_v^n \xi) - E(P^n) < 0 \quad (2)$$

Forward test vector divergence given as follows;

$$P^{n+1} = P^n + \Psi_v^n \xi \quad (3)$$

Backward test is realized according to following formulation:

$$E(P^n - \Psi_v^n \xi) - E(P^n) < 0 \quad (4)$$

Backward test vector divergence given as follows;

$$P^{n+1} = P^n - \Psi_v^n \xi \quad (5)$$

Basic steps of MO-SMDO algorithm can be summarized as,

Step 1: Set an initial value to the parameter vector $P$ and calculate objective function $E(P^0)$.

Step 2: For all member of the parameter vector $P^n$, apply forward test by Eq. (2) and for backwards test by Eq. (4). When, any component satisfies $J(P_i^{n+1}) - J(P_i^n) < 0$ condition, update parameter vector $P^n$.

Step 3: Calculate the multi objective function $E = \min(w_1 E_{TSH} + w_2 E_{e,\text{sec},out})$, where $w_1$ and $w_2$ are weighting functions. Weighting function is used to adjust the related error function values in the same level.

Step 4: if $f(P^n) < E$ stop searching. Otherwise increase the iteration number ($n = n+1$) and go to Step 2.

## 4. SIMULATION RESULT

In this section, simulation results are presented, and performances of PI controllers are compared. Simulation model of Refrigeration Systems based on Vapour Compression in Fig.4 is used as a plant in simulation study, for $T_{e,\text{sec},out}$ and $TSH$ respectively. This Matlab simulation model is run in 1200 seconds with Matlab solver "ode23tb". For each number of the optimization process, corresponding simulation model is run under the same conditions. All results are compared according to this model.

PI parameters of the system are optimized then the existing and new results are compared. Proposed optimization algorithm runs 100 iterations due to have some time limitation These figures show that related optimization algorithm works well. Decrease of cost function values indicates the improvement of control system response according to multi objective function in Eq. 1. Fig. 5 compares the performance of two PI controller responses for $T_{e,\text{sec},out}$. Existing PI controller parameters are given in (Bejarano et al., 2018)) (blue line in the figure). Proposed PI controller parameters (red line in the figure) are optimized according to multi objective function in Eq.1 with conditional integral (CI).

As seen in figures 5, 6 and 7, optimized multi objective PI controller with CI gives better set-point, settling time and less steady state error performance compared to existing PI control structure for $T_{e,\text{sec},out}$. Proposed PI controller, optimized with multi objective function in Eq. 1, can significantly reduce steady state error. Consequently, proposed PI control parameters can present better control performance according to the existing PI controller. At the end of the optimization, PI controller parameters are optimized as $k_p = 1.5027$ $k_i = 2.1179$. In the optimization process discreet PI controller is obtained directly and apply to simulation model.

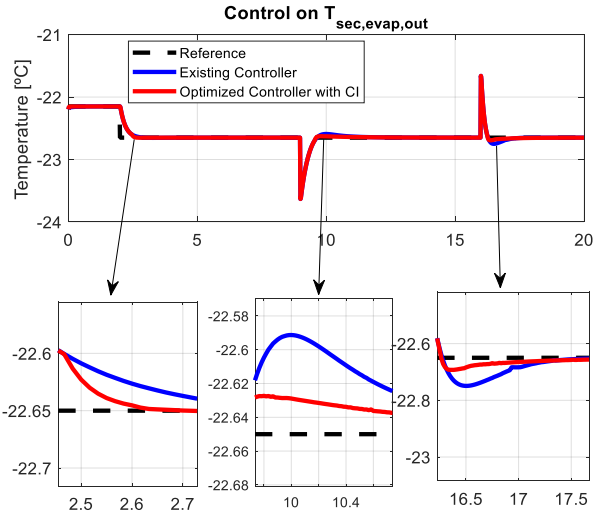

Fig.5. Response signal of $T_{e,\text{sec},out}$

Fig. 6 compares results between optimized PI controller with existing PI controller response for TSH. As seen in the figure, optimized multi objective PI controller with CI gives better set-point, settling time and less steady state error performance compared to existing PI control structure for *TSH* like in Fig. 5. On the other hand, using of the proposed PI controller, optimized with multi objective function, can significantly reduce steady state error. Consequently, proposed PI control parameters and structure can present better control performance than existing PI controller.

The control signal of the Refrigeration Systems based on Vapour Compression in Fig. 7 shows that the control response is saturated for optimized PI and existing controller for compressor speed. Nevertheless, optimized PI parameters provide better results. On the other hand, corresponding controller index values given in Table 1 show that the optimized PI controller structure with CI gives the least value. This value validates the proposed method has better control performance than the existing control structure. Furthermore, this system takes too much time for the optimization process. If the simulation system is run for more than 100 iterations. J can be better than previously optimization cycle.

5. CONLUSION

The paper presents PI controller tuning method for Refrigeration Systems based on Vapour Compression according to MO-SMDO methods with conditional integral. Test results validate that the optimized PI controller's parameters can yield better control performance than the existing control structure. This study also shows that the usage of the conditional integral structure is very useful for reduction of the steady state error. Thus, related index values (J) in Table 1 are compared for optimized PI controller's parameters with and without using conditional integral to show priority of the proposed method.

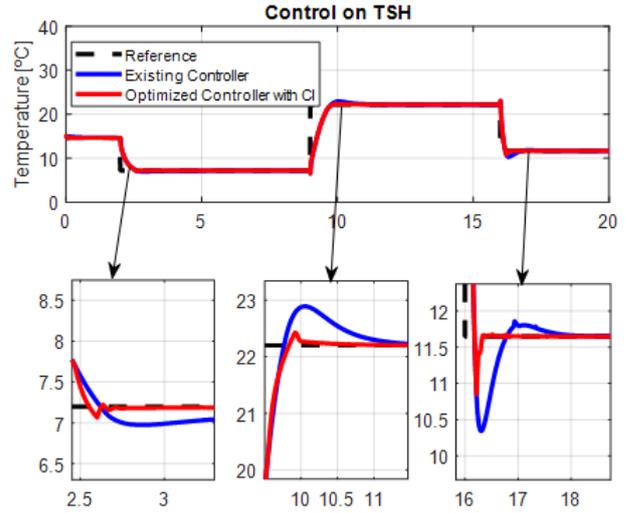

Fig.6. Response signal of *TSH*

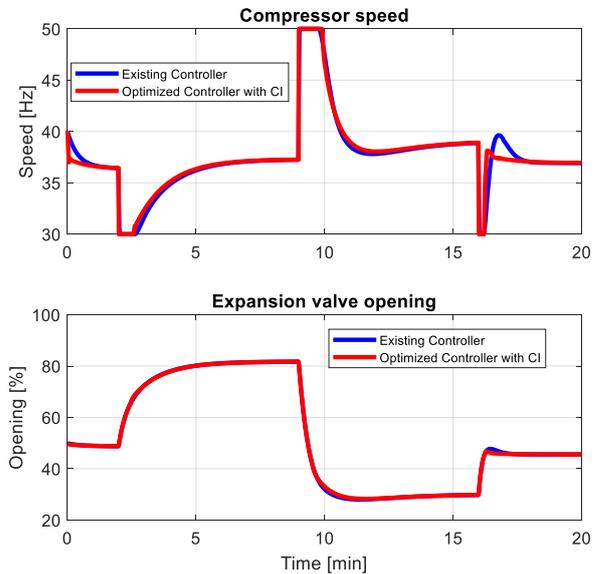

Fig.7. Control Signal of the system

Table 1. Comparisons of the J index

|  | $J(c_1, c_2)$ |
|---|---|
| **Existing MIMO PID C2** | 0.68209 |
| **Optimized PI** | 0.77 |
| **Optimized PI with CI** | **0.6532** |

# ACKNOWLEDGEMENT

This study is supported by The Scientific and Technological Research Council of Turkey (TUBITAK-BIDEP) with 2214/A program number.